\documentclass[12pt]{iopart}
 \usepackage{iopams}
\usepackage{graphicx}
\graphicspath{{./graphics/}}
\usepackage{color}
\usepackage{amsfonts}
\usepackage{cite}
\usepackage[colorlinks=true,
            linkcolor = black,
            urlcolor  = blue,
            citecolor = black,
            anchorcolor = blue]{hyperref}
\newcommand{\dd}{\text{d}}      

\begin{document}

\title{Optimized finite-time information machine}
\author{Michael Bauer, Andre C. Barato and Udo Seifert}

\address{
{II.} Institut f\"ur Theoretische Physik, Universit\"at Stuttgart,
  70550 Stuttgart, Germany}

\begin{abstract}
We analyze a periodic optimal finite-time two-state information-driven machine that extracts work from a single heat bath exploring imperfect measurements. 
Two models are considered, a memory-less one that ignores past measurements and an optimized model for which the feedback scheme consists of 
a protocol depending on the whole history of measurements. Depending on the precision of the measurement and on the period length the optimized model 
displays a phase transition to a phase where measurements are judged as non-reliable.
We obtain the critical line exactly and show that the optimized model leads to more work extraction in comparison to the memory-less
model, with the gain parameter being larger in the region where the frequency of non-reliable measurements is higher. 
We also demonstrate that the model has two second law inequalities, with the extracted work being bounded by the change of the entropy of the system  and 
by the mutual information.

\end{abstract}


\def\i{{\scriptstyle {\cal I }}} 
\def\I{{\cal I}}
\def\dd{\mathrm d}
\def\plog{\mathrm{plog}}


\section{Introduction}

The thermodynamics of information processing is a very active area of research. Whereas central concepts in this field have been developed a while ago \cite{szil29,land61,benn82}, more recently the fluctuation relation obtained 
by Sagawa and Ueda \cite{saga10} has shown that stochastic thermodynamics \cite{seif12} provides a convenient framework to study the relation between information and thermodynamics.
Moreover, ingenious experiments with small systems \cite{toya10a,beru12} verifying second law inequalities that involve information have played an important role in triggering the recent avalanche of papers. 
These works deal with the derivation of fluctuation relations and second law inequalities \cite{touc00,touc04,cao09, saga10,ponm10,horo10,abre11a,kund12,saga12,saga12b,ito13,alla09,hart14} and 
the study of specific models \cite{cao04,horo11,horo11a,abre11,gran11,kish12,espo12b,stra13,horo13,gran13}.
 
In  finite-time thermodynamics the issue of optimal protocols is of central importance. A recent result within
stochastic thermodynamics has been the observation that the optimal protocol has discontinuities at the beginning and end of 
the finite-time process \cite{schm07,then08,schm08,gome08,espo10a,aure11,aure12}. In  information processing, optimal protocols have so far been analyzed 
for the maximal work extraction in a feedback driven system described by an one-dimensional over-damped Langevin equation  \cite{baue12} and for the minimum dissipated heat
in an erasure process \cite{dian13a,zulk14}. 

In this paper we study a paradigmatic discrete two-state model \cite{horo13,espo10a,kuma11a,bara14}, where the work extraction,
performed by lifting and lowering one energy level, is driven by feedback. Besides applying the optimal protocol leading to 
the maximal work during one period, this information machine will also be optimized in the sense that the protocol takes the whole history of measurements into account.

We show that this optimized feedback strategy leads to more work extraction in comparison to a simple memory-less machine. Moreover,
we observe a phase transition, where in one phase the machine always lifts the state indicated by the last measurement as empty and in the other phase the 
state measured as occupied is lifted with a certain frequency. The extracted work is observed to be bounded by two quantities: the familiar mutual information between system and controller
and the change of the entropy of the system. While the second bound is valid for every measurement trajectory, the first becomes valid only after an average over measurement trajectories is taken.
Finally, we show that the memory-less model allows for a different physical interpretation of the system interacting with a tape, i.e., a sequence of bits. This memory-less model then corresponds to
a generalization of the model recently introduced in \cite{mand12} (see also \cite{bara14,mand13,bara13,deff13}).

The paper is organized as follows. In Sec. \ref{sec2} we obtain the optimal protocol for a single period. The full feedback driven models are defined in Sec. \ref{sec3}.
The phase transition and gain parameter for the optimized model are studied in Sec. \ref{sec4}. In Sec. \ref{sec5} the different second law inequalities valid for the model are analyzed.
We conclude in Sec. \ref{sec6}.

\section{Two-state finite-time process}
\label{sec2}

\subsection{Model}

\begin{figure}
\centering\includegraphics{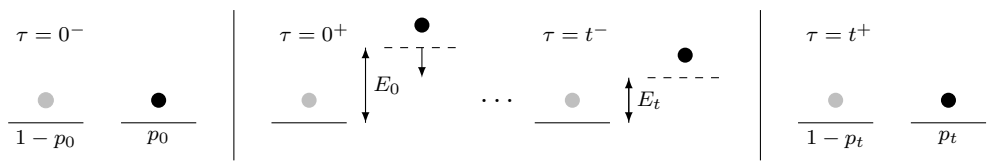}
\caption{Representation of the finite-time process. Initially, at $\tau=0^-$, the entropy of the system is $H(p_0)=-p_0\ln p_0 -(1-p_0)\ln(1-p_0)$. At $\tau=0^+$ the level with lower probability $p_0\le 1/2$ is
lifted. For $0\le\tau\le t$ this energy level is lowered with protocol $E_\tau$. At time $\tau=t$ this energy level is set from $E_t$ back to $0$ extracting work $E_t$ if this level had been occupied at $\tau=t^-$.}
\label{fig1}
\end{figure}

The model analyzed in this paper is a two-state system where the time dependent energy of the upper level $E_\tau\ge 0$ can be controlled by an external protocol in the time interval $0\le \tau \le t$. The lower level has always energy zero.
This system is connected to a heat bath at temperature $T$ and a work reservoir.  We consider a finite-time process with duration $t$, where both energy levels are zero immediately before starting $\tau=0^-$ 
and immediately after finishing $\tau=t^+$, see Fig. \ref{fig1}.  These initial and final jumps of $E_\tau$ are a generic feature of optimal  protocols \cite{schm07}.

Denoting the occupation probability of the upper level by $p_\tau$, the time derivative of the average internal energy reads 
\begin{equation}
\frac{d}{d\tau}E_\tau p_\tau= \dot{E}_\tau p_\tau+E_\tau\dot{p}_\tau,
\end{equation}
where the dot represents a time derivative throughout the paper.
This is the first law of thermodynamics, where $\dot{w}= -\dot{E}_\tau p_\tau$ is identified as the rate of extracted work and $\dot{q}_\tau= E_\tau\dot{p}_\tau$ as the rate of absorbed heat.
This identification means that if a jump occurs heat is exchanged with the heat bath and if the energy level changes work is exchanged with the work reservoir.
The extracted work in the time interval $t$ then becomes
\begin{equation}
W_t= -\int_0^t \dot{E}_\tau p_\tau \dd \tau  + E_tp_t-E_0p_0,
\end{equation}
where the boundary terms comes from the discontinuities in $E_\tau$ represented in Fig. \ref{fig1}.
Since the variation of the internal energy is zero, the extracted work equals the heat absorbed from the heat bath, i.e.,
\begin{equation}
 W_t= Q_t=  \int_0^t E_\tau\dot p_\tau \dd \tau.
\label{eq:work_in}
\end{equation}
Even though the system is connected to a single heat bath and the 
variation of the internal energy is zero, it is still possible to extract work due to the  increase in the entropy of the system.
More precisely, the second law for such isothermal process establishes that the extracted work 
is bounded by the change of the entropy of the system, i.e.,
\begin{equation} 
W_t\le k_B T [H(p_t)-H(p_0)],
\label{secondH}
\end{equation}
where $H(p)\equiv-p\ln p -(1-p)\ln(1-p)$. In this paper we set $k_BT=1$ and, in order to have work extraction, we restrict 
to the case $p_0\le p_t\le 1/2$.

\subsection{Optimal Protocol}

The optimal protocol $E_\tau$ that leads to the maximal work extraction for given time interval $t$ and initial occupation probability $p_0$ is calculated in the remaining of this section. The master equation reads 
\begin{equation}
\dot p_\tau=-\omega_-p_\tau+\omega_+[1-p_\tau],
\label{eq:masterequation}
\end{equation}
where $\omega_+$ ($\omega_-$) is the time dependent transition rate to (from) the upper level. These rates must fulfill the detailed balance relation $\omega_{-}/\omega_{+}=e^{E_\tau}$.
For convenience, we choose
\begin{equation}
\omega_+=e^{-E_\tau}\qquad\textrm{and}\qquad \omega_-=1.
\label{eq:choice_of_rates_1}
\end{equation}
Following the analysis for a symmetric choice of rates \cite{espo10a}, the optimal protocol and the corresponding maximal extracted work is found by considering the Lagrangian
\begin{equation}
 L(p,\dot p)\equiv\dot p \ln \left( \frac{1-p}{p+\dot p} \right),
\end{equation}
where the work is then written as  
\begin{equation}
 W_t=\int_0^t   L(p,\dot p) \dd \tau.
\label{eq:work_int}
\end{equation}
Since $L(p,\dot p)$ does not  explicitly depend on $\tau$, we have the following constant of motion, 
\begin{equation}
 K\equiv L-\dot p \frac{\partial L}{\partial \dot p}=\frac{\dot p^2}{p+\dot p}\ge0.
\label{eq:K_def}
\end{equation}
Introducing  the variable $r_\tau\equiv p_\tau/K\ge0$, equation (\ref{eq:K_def}) becomes  
\begin{equation}
\dot{r}_\tau=\frac{1}{2} + \frac{1}{2}\sqrt{1+4r_\tau}\ge1.
\end{equation}
The solution of this equation is 
\begin{equation}
 \dot r_\tau =-\frac 1 2 \plog_{-1}(-2 \dot r_0 e^{-2\dot r_0} e^{-\tau})\equiv f_\tau(\dot r_0).
\label{eq:dglq}
\end{equation}
where $\plog_{-1}(x)$ is the lower branch of product logarithm \cite{corl96}. Using relations $r_\tau= \dot r_\tau^2-\dot r_\tau$, 
$K=p_0/r_0=p_0/[\dot r^2_0-\dot r_0]$, and (\ref{eq:dglq}), the extracted work (\ref{eq:work_int}) becomes a function of the single variable $\dot r_0$. The maximal work is then obtained 
for $\dot r_0=a$, where $\frac{d}{d \dot r_0}W|_{\dot r_0= a}=0$. In this way, $a(p_0)$ is given by the solution of the transcendental equation 
\begin{equation}
f^2_t(a) \left(\exp[{1/f_t(a)}]+1\right)-f_t(a)=\frac{a(a-1)}{p_0}.
\label{transc}
\end{equation}

\begin{figure}
\centering\includegraphics[width=0.99\textwidth]{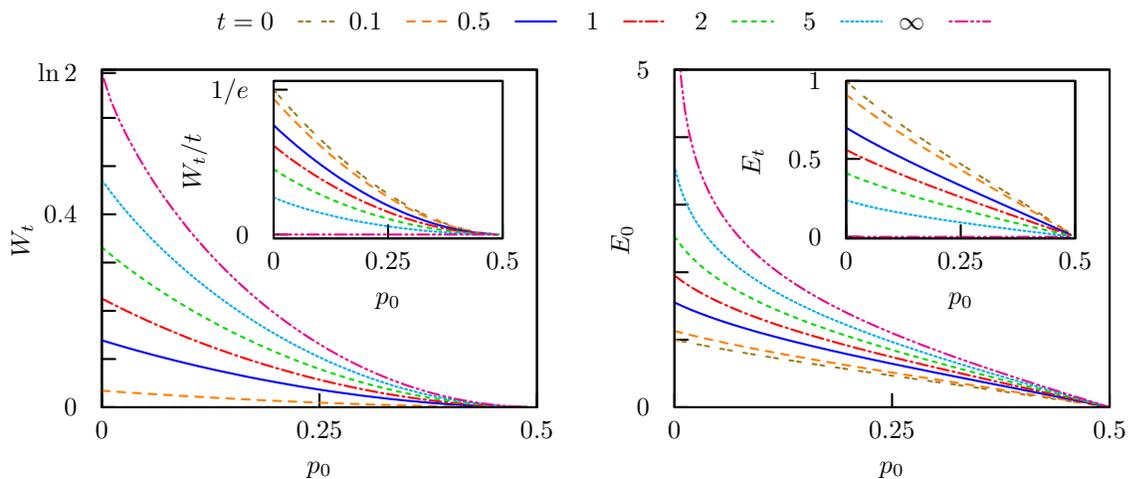}
\caption{Maximal work $W_t(p_0)$ for different values of $t$ on the left, with the power $W_t(p_0)/t$
plotted in the inset. The initial energy jump $E_0(p_0)$ and the final energy jump $E_t(p_0)$ in the inset are plotted on the right.}
\label{fig2}
\end{figure}

For convenience the optimal protocol and corresponding maximal work are simply denoted by  $E_\tau(p_0)$ and $W_t(p_0)$, respectively.
From (\ref{eq:work_int}) the maximal work that can be extracted for fixed $t$ and $p_0$ is  
\begin{equation}
 \hspace{-1.5cm}W_t(p_0)=-\ln\left( \frac{1-\frac{p_0}{a^2-a}[f^2_t(a)-f_t(a)]}{1-p_0}\right)+p_0\left[\ln\left(\frac{p_0a}{(1-p_0)(a-1)}\right) +a^{-1} \right],
\label{maxwork}
\end{equation}
and the corresponding optimal protocol reads   
\begin{equation}
E_\tau(p_0)=  \ln\left(\frac{p_0f_\tau(a)+a^2-a}{p_0f^2_\tau(a)}-1\right),
\label{protopt}
\end{equation}
where $a(p_0)$ is given by the solution of equation (\ref{transc}). In Fig. \ref{fig2}, we plot the maximal work, the power and the discontinuities of the optimal protocol as a function of $p_0$ for given $t$.
The optimal work is a decreasing function of $p_0$, with full knowledge of the initial state leading to the maximal work extraction for fixed $t$. 
By increasing $t$, the work increases  whereas the power $W_t(p_0)/t$ decreases, going to zero in the limit 
$t\to \infty$. The initial and final energy jumps decrease with $p_0$, being maximal for $p_0=0$. The initial jump  $E_0(p_0)$ increases with $t$, while the final jump $E_t(p_0)$ decreases with $t$.
More precisely, for $t\to 0$ we have $E_0(p_0)=E_t(p_0)$ and the difference between the jumps grows with $t$, with $E_0(p_0)$ reaching its maximal value and $E_t(p_0)\to 0$ for $t\to \infty$. 

Finally it is useful for the following discussion to give $p_t$ for the optimal protocol explicitly as  
\begin{equation}
p_t= \frac{p_0}{a^2-a}[f^2_t(a)-f_t(a)].
\label{pt}
\end{equation}

\section{Feedback driven machine}
\label{sec3}
\subsection{Imperfect measurements}
An information driven machine periodically repeats the process explained in the previous section using feedback control. Measurements and feedback drive the
work extraction by resetting  the entropy of the system at the end of the time interval. We denote the state of the system just before starting period $i$  by $x_i$  and the measurement 
by $m_i$, where $x_i= -1$ ($x_i=+1$) means that the left (right) state is occupied while $m_i= -1$ ($m_i=+1$) corresponds to measuring the left (right) state as being occupied. 
The conditional probability of the measurement is defined as 
\begin{equation}
P(m_i|x_i)\equiv \left\{
\begin{array}{ll} 
 1-\epsilon & \quad \textrm{if $m_i=x_i$}, \\
 \epsilon & \quad \textrm{if $m_i\neq x_i$}, 
\end{array}\right.\,
\label{error}
\end{equation}
where $\epsilon$ is the measurement error. The machine never knows the real state of the system $x_i$ and has access only to the history of measurements $m_1^i=\{m_1,m_2,\ldots,m_i\} $.
Hence, in all calculations that follow the state of the system is always averaged out.


\subsection{Optimized machine}

First we consider a feedback procedure with a protocol taking the whole measurement trajectory $m_1^i=\{m_1,m_2,\ldots,m_i\} $ into account.
We are interested in the probability of being at state $x_i$ given the history of measurements $m_1^i$, which is denoted $P(x_i|m_1^i)$.
For this feedback scheme the initial occupation probability of the level that will be raised at the beginning of period $i$ is 
\begin{equation}
p_0^{(i)}= \textrm{min}\{P(x_i=m_i|m_1^{i}),P(x_i=-m_i|m_1^{i})\}\leq \frac{1}{2}.
\label{defp0i}
\end{equation}
This means that contrary to the naive procedure of lifting the level $-m_i$ independent of the measurement history it is also possible to make the unusual choice of lifting the level $m_i$. In this second case, 
the level indicated by the last measurement as occupied is lifted: the measurement is judged to be not reliable. Moreover, the machine applies 
the protocol $E_\tau(p_0^{(i)})$, which takes into account the whole history of measurements by using the history dependent initial probability $p_0^{(i)}$.

In \ref{appa} we show that the  initial probability $p_0^{(i)}$ fulfills a nonlinear recursion relation.   
Denoting  by $p_t^{(i-1)}$ the probability at the end of the period $i-1$ that  is obtained from the function (\ref{pt}) with initial probability $p_0^{(i-1)}$, we define the functions  
\begin{equation}
 F_+(p_0^{(i-1)})\equiv \frac{\epsilon p_t^{(i-1)}}{1-q_t^{(i-1)}} 
\end{equation}
and
\begin{equation}
F_-(p_0^{(i-1)})\equiv \frac{\epsilon(1-p_t^{(i-1)})}{q_t^{(i-1)}}, 
\end{equation}
where 
\begin{equation}
q_t^{(i-1)}\equiv p_t^{(i-1)}+\epsilon(1-2p_t^{(i-1)}).
\end{equation}
The recursion relation for $p_0^{(i)}$  then reads
\begin{equation}
p_0^{(i)}=  \left\{
\begin{array}{ll} 
 F_+(p_0^{(i-1)}) & \quad \textrm{if $\tilde{z}_{i}=1$}, \\
 \textrm{min}\{F_-(p_0^{(i-1)}),1-F_-(p_0^{(i-1)})\} & \quad \textrm{if $\tilde{z}_{i}=-1$}.
\end{array}\right.\,
\label{recursion}
\end{equation}
As explained in \ref{appa}, the variable $\tilde{z}_i$ has the purpose of identifying  whether the measurement outcome $m_i$ corresponds to
the upper or the lower level of the interval $i-1$, with $\tilde{z}_i=-1$ if the upper level is $m_i$   and $\tilde{z}_i=1$ if the lower level is $m_i$.
We call this machine taking the history of measurements $m_1^i$ into account the optimized machine because, as we will see in Sec. \ref{sec4}, it leads
to more work extraction then a simple memory-less machine which we define next.

\subsection{Memory-less machine}

A memory-less feedback scheme that only takes the last measurement into account would be to simply apply a protocol for which the level raised for the next period is just the state measured as empty. Hence,
for a measurement outcome $m_i$, the level $-m_i$ is lifted at the beginning of period $i$. As we show in \ref{appb}, where the memory-less machine is
more explicitly defined, the average initial occupation probability of the upper level is $\epsilon$, independent of the protocol. 
Therefore, the appropriate choice for a protocol that must be independent from the whole measurement history and corresponds to the memory-less version of the optimized machine is $E_\tau(\epsilon)$, 
which is obtained from (\ref{protopt}) with $p_0=\epsilon$.

\section{Gain and phase transition}
\label{sec4}

The work extracted during period $i$ with the optimized  machine is denoted by $W_t^{(i)}=W_t(p_0^{(i)})$.
For a given measurement realization $m_1^N$ we define 
\begin{equation}
\overline{W}_t\equiv \frac{1}{N}\sum_{i=1}^{N}W_t^{(i)}.
\label{avg}
\end{equation}
The average work $\langle \overline{W}_t \rangle$ is obtained by considering the limit $N\to \infty$ and averaging over all measurement trajectories, where the brackets denote
this average over measurement trajectories. Numerical simulations for large enough $N$ indicate that 
$\overline{W}_t$ (and other observables we calculate below) is independent of the numerically generated measurement history, 
i.e., self-averaging. Therefore, we calculate the average work by generating a single long measurement history.

\begin{figure}
\centering\includegraphics{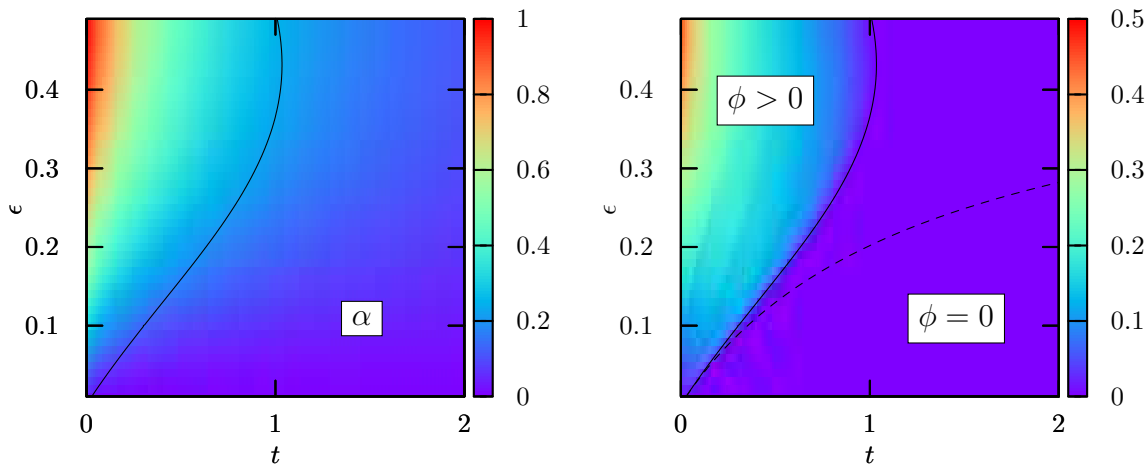}
\caption{The gain parameter $\alpha$ (left) and the order parameter $\phi$ (right) as functions of the time interval $t$ and the  measurement error $\epsilon$. The results are obtained by
numerically generating a measurement trajectory of length $10^7$. The full black critical line is obtained analytically from (\ref{eqcrit}) and the dotted line on the right panel from (\ref{eqdot}).}
\label{fig3}
\end{figure}

For the memory-less machine the average work is just $W_t(\epsilon)$, as demonstrated in \ref{appb}. The improvement of the optimized in relation to the memory-less machine is quantified by the gain parameter 
\begin{equation}
\alpha\equiv 1-\frac{W_t(\epsilon)}{\langle \overline{W}_t \rangle}.
\end{equation} 
Naively one expects the optimized machine that takes the history of measurements into account to extract more work than the simple machine. This
expectation is confirmed by numerical simulations, from which we observe that $\alpha\ge 0$. For $\alpha= 0$ the work extraction in the memory-less model would be the same as in the optimized model
and for $\alpha\to 1$ the work extraction is much higher with the optimized model. In Fig. \ref{fig3}, we plot $\alpha$ in the $(t,\epsilon)$-plane. 
The gain approaches $1$ for small $t$ and $\epsilon$ close to $1/2$, where non-reliable measurements are more likely to occur.

It turns out that the optimized model displays a phase transition. The order parameter for this transition $\phi$ is the frequency at which the state $m_i$ is lifted,
i.e.,
\begin{equation}
\phi\equiv  \left\langle \frac{1}{2N}\sum_{i=1}^{N}(1-\sigma_i)\right\rangle,
\end{equation}  
where $\sigma_i=-1$ if the measurement is not reliable and $\sigma_i=1$ if the measurement is reliable (see \ref{appa} for a precise definition).
The numerical calculation of this order parameter is also shown in Fig. \ref{fig3}. We can clearly see a phase transition with $\phi= 0$ below a critical threshold $\epsilon_c(t)$.
Numerics indicates a second order phase transition.

The optimized machine has two advantages in relation to the memory-less machine: it lifts the level $m_i$ if the last measurement is not reliable and it uses
a history dependent protocol $E_\tau(p_0^{(i)})$. By comparing $\alpha$ with $\phi$ in Fig. \ref{fig3}, we see that in the phase $\phi>0$ there is a substantial 
gain. This means that the first advantage is the key feature leading to more work extraction for the optimized feedback scheme. Moreover, as shown  
in \ref{appa}, in the phase $\phi=0$ the average initial occupation probability is $\epsilon$. Hence, $\alpha>0$ in this phase arises 
from the fact that the function $W_t(p_0)$ plotted in Fig. \ref{fig2} is convex, implying  $\langle W_t^{(i)}\rangle_i\ge W_t(\epsilon)$, where the average $\langle.\rangle_i$ is
defined in (\ref{defavgi}).
   
As we show in \ref{appc}, the critical line $\epsilon_c(t)$ can be obtained analytically from the transcendental equation (\ref{eqcrit}). It is in perfect agreement with numerical results, 
as shown in Fig. \ref{fig3}.

\section{Second law inequalities}
\label{sec5}

\subsection{Efficiency and power}

The second law for feedback driven systems \cite{cao09} states that the average extracted work is bounded by the average mutual information between system and controller due to measurements. 
The mutual information between the system and the controller due to the measurement $m_i$ is defined as 
\begin{eqnarray}
I_t^{(i)}&\equiv \sum_{m_{i},x_{i}} P(m_{i},x_{i}|m_1^{i-1})\ln\frac{P(m_{i},x_{i}|m_1^{i-1})}{P(m_{i}|m_1^{i-1})P(x_{i}|m_1^{i-1})}\nonumber\\
&= H(q_t^{(i-1)})-H(\epsilon).
\label{mutuopt}
\end{eqnarray}
We denote the average mutual information by $\langle\overline{I}_t\rangle$, so that  the efficiency of the optimal machine reads
\begin{equation}
\eta\equiv \frac{\langle\overline{W}_t\rangle}{\langle\overline{I}_t\rangle}.
\end{equation}
\begin{figure}
\centering\includegraphics{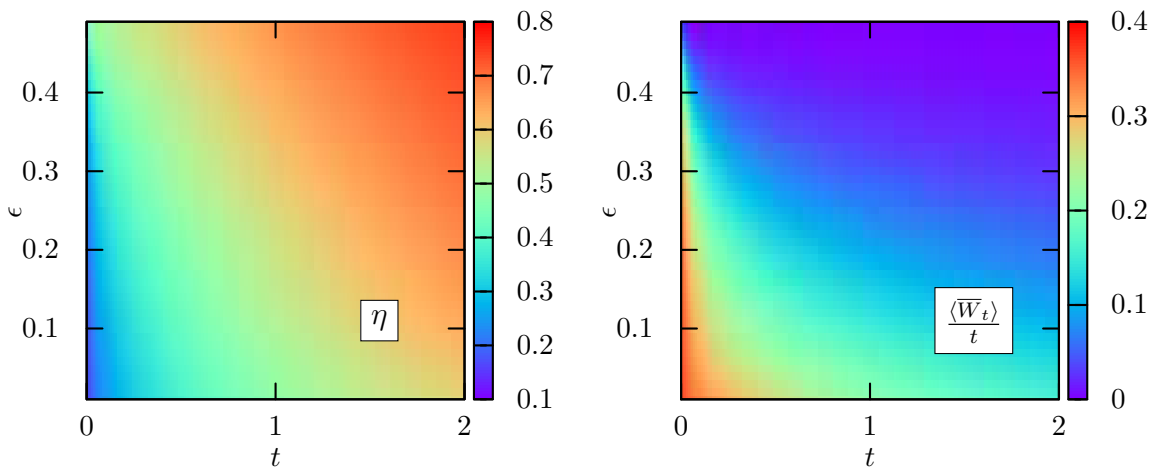}
\caption{The efficiency $\eta$ and the average power $\langle\overline{W}_t\rangle/t$ as functions of the time interval $t$ and the  measurement error $\epsilon$. The results are obtained by
numerically generating a measurement trajectory of length $10^7$.}
\label{fig6}
\end{figure}

In Fig. \ref{fig6} we show the numerically calculated efficiency $\eta$ and power $\langle\overline{W}_t\rangle/t$ for the optimized model in the $(t,\epsilon)$-plane. 
Increasing the time period $t$ increases the efficiency but decreases the power of the machine. For fixed $t$, the efficiency increases for increasing measurement error $\epsilon$.
Hence, maximum power is obtained for small $\epsilon$ and small $t$, which is, however, a rather inefficient case with $\eta\lesssim0.2$.
 
\subsection{Two second law inequalities }

Another bound on the extracted work is provided by the Shannon entropy change, as expressed in (\ref{secondH}), which for the optimized model can be written as
\begin{equation}
W_t^{(i)}\le \Delta H_t^{(i)},
\label{boundW}
\end{equation}
where the Shannon entropy change in interval $i$ is  
\begin{equation}
\Delta H_t^{(i)}\equiv H(p_t^{(i)})-H(p_0^{(i)}). 
\end{equation}
The inequality (\ref{boundW}) is then valid for a fixed measurement trajectory, whereas the standard second law for feedback driven systems $\langle\overline{W}_t\rangle\le \langle\overline{I}_t\rangle$ is valid 
only after an average over measurement trajectories is taken. By numerical inspection we observe that $I_t^{(i)}$ (or $I_t^{(i+1)}$ ) can be smaller than $W_t^{(i)}$.
Furthermore, by taking the average for large $N$ we find $\langle \overline{I}_t\rangle= \langle \overline{\Delta H}_t\rangle$ within numerical errors.
This equality can be demonstrated with the following heuristic argument. Rearranging the terms in the mutual information we get
\begin{eqnarray}
I_t^{(i+1)} & = H(q_t^{(i)})- H(\epsilon)\nonumber\\
& =H(p_t^{(i)})-q_t^{(i)}H\left(\frac{\epsilon(1-p_t^{(i)})}{q_t^{(i)}}\right)-(1-q_t^{(i)})H\left(\frac{\epsilon p_t^{(i)}}{1-q_t^{(i)}}\right)\nonumber\\
& =  H(p_t^{(i)})-\langle H(p_0^{(i+1)})\rangle_{i+1}.
\label{IeDH}
\end{eqnarray}
where the average $\langle.\rangle_{i+1}$ is defined in (\ref{defavgi}). From this equation it is clear that the average mutual 
information and the average Shannon entropy change differ only by boundary terms, which for large $N$ should be irrelevant, implying
$\langle \overline{I}_t\rangle= \langle \overline{\Delta H}_t\rangle$. 

\subsection{Memory-less machine as a system interacting with a tape}

As in \ref{appb} we now consider a memory-less machine using an arbitrary protocol $\tilde{E}_\tau$. Equation (\ref{IeDH}) is also valid for the memory-less case and, therefore, the average Shannon entropy change should be equal to the
average mutual information for large $N$. This result was confirmed numerically for the protocol $\tilde{E}_\tau=E_\tau(\epsilon)$ and for $\tilde{E}_\tau=E$, which corresponds to
the energy level held fixed during the whole time interval. Similar to (\ref{mutuopt}), the mutual information depending on $m_1^{i-1}$ reads 
\begin{eqnarray}
\tilde{I}_t^{(i)}&= H(\tilde{q}_t^{(i-1)})-H(\epsilon),
\end{eqnarray}
where $\tilde{q}_t^{(i-1)}$ is defined in (\ref{qtilde}). Denoting by $\tilde{p}_t$ the solution of the master equation (\ref{eq:masterequation}) with protocol $\tilde{E}_\tau$ and initial probability $\epsilon$
we define 
\begin{equation}
\tilde{I}_t\equiv H(\tilde{q}_t)-H(\epsilon),
\end{equation}
where $\tilde{q}_t= \tilde{p}_t+\epsilon(1-2\tilde{p}_t)$. Since $\tilde{q}_t^{(i)}$ (and $\tilde{p}_t^{(i)}$) is linear in $\tilde{p}_0^{(i)}$, it follows that $\langle\tilde{q}_t^{(i)}\rangle_i=\tilde{q}_t$, where the average $\langle .\rangle_i$ 
is defined in (\ref{avgdum}). From the fact that the Shannon entropy is concave we obtain that $\tilde{I}_t$ provides an upper bound on the average mutual information
\begin{equation}
\langle\tilde{I}_t\rangle\equiv \left\langle\frac{1}{N} \sum_{i=1}^{N}\tilde{I}_t^{(i)}\right\rangle.
\end{equation}

As the average extracted work is equal to the work extracted in the first period (see \ref{appb}), if before the first measurement both states are equally probable, 
the average extracted work is also bounded by the Shannon entropy change in the first period
\begin{equation}
\Delta\tilde{H}_t\equiv H(\tilde{p}_t)-H(\epsilon).
\end{equation}
Comparing with the other bounds we have $\Delta\tilde{H}_t\le \tilde{I}_t$ and, numerically, for the protocols $\tilde{E}_\tau=E_\tau(\epsilon)$ and $\tilde{E}_\tau=E$, we
observe  $\Delta\tilde{H}_t\le \langle \tilde{I}_t\rangle$. We conjecture that this entropy change provides the strongest bound on the extracted work.

The inequality 
\begin{equation}
\tilde{W}_t(\epsilon)\le \Delta\tilde{H}_t
\label{inegen}
\end{equation}
for the protocol $\tilde{E}_\tau=E$ has been recently studied in \cite{bara14}. In this reference it has been shown that the two-state model
can also be interpreted as a tape, i.e., a sequence of bits, interacting with a thermodynamic system. In this interpretation the entropy change
is dumped to a tape or information reservoir. The inequality (\ref{inegen}) means that the extracted  work is bounded by the Shannon entropy change of the tape, 
which is initially $H(\epsilon)$ and becomes $H(p_t)$ after the system has written information to it. This model for a tape interacting with a thermodynamic 
system has been introduced by Mandal and Jarzynski \cite{mand12}, for a model with six
instead of two states and a protocol that is also held fixed during the whole time interval. By showing that inequality (\ref{inegen}) is valid
also for arbitrary $\tilde{E}_\tau$ protocols, we thus obtain that their model can be generalized to arbitrary time-dependent protocols.

\section{Conclusion}
\label{sec6}

We have studied a two-state finite-time optimized information-driven machine. Besides utilizing the optimal protocol, this machine is also optimized
in the sense that the feedback scheme takes into account the whole history of measurements. We have shown that the optimized machine leads to more work extraction in comparison  to a simple
memory-less machine that does not take the full measurement trajectory into account. 

This optimized model displays a phase transition with the frequency at which non-reliable measurements occur being the  order parameter. In the region of the phase diagram  where non-reliable measurements 
occur with a higher frequency the gain parameter, characterizing the improvement of the optimized in relation to the memory-less machine, was found to be high. Hence the possibility of lifting 
the state last measured as occupied if the measurement is non-reliable is the key feature that makes the optimized model perform better. Moreover, analyzing the
recursion relations for the initial occupation probability of the upper level we have obtained the critical line exactly.

We have shown that the work extraction is bounded both by the Shannon entropy change and the mutual information. While the first bound is valid for every measurement trajectory the second
is valid only after averaging over the measurements. In this case, both bounds become the same. Moreover, for the memory-less model we have demonstrated that the average extracted work is bounded by 
the Shannon entropy change of the first period. This inequality allows for an interpretation of the model as a thermodynamic system interacting with a tape, 
thus generalizing the model introduced in \cite{mand12}.

\appendix

\section{Iterative relation for the optimized model}
\label{appa}
In this appendix, we obtain the nonlinear recursive relation for the initial occupation probability of the upper level of the optimized machine (\ref{recursion}).
From relations 
\begin{equation}
P(m_i|m_1^{i-1})=\sum_{x_i} P(x_i,m_i|m_1^{i-1})
\end{equation}
and
\begin{equation}
P(x_i,m_i|m_1^{i-1})= P(m_i|x_i,m_1^{i-1})P(x_i|m_1^{i-1})= P(m_i|x_i)P(x_i|m_1^{i-1}),  
\end{equation}
we obtain
\begin{eqnarray}
P(m_i|m_1^{i-1}) & = (1-\epsilon)P(x_i=m_i|m_1^{i-1})+\epsilon P(x_i=-m_i|m_1^{i-1})\nonumber\\
& =P(x_i=m_i|m_1^{i-1})+\epsilon[1-2P(x_i=m_i|m_1^{i-1})],
\label{errorh}
\end{eqnarray}
where we used the definition of measurement error (\ref{error}).
Using the relation 
\begin{equation}
P(x_i|m_1^{i})= P(x_i,m_i|m_1^{i-1})/P(m_i|m_1^{i-1})
\end{equation}
the conditional probability $P(x_i|m_1^i)$ can then be written as
\begin{equation}
P(x_i|m_1^{i})= \left\{
\begin{array}{ll} 
 \frac{(1-\epsilon)P(x_i=m_i|m_1^{i-1})}{P(x_i=m_i|m_1^{i-1})+\epsilon[1-2P(x_i=m_i|m_1^{i-1})]} & \quad \textrm{if $x_i=m_i$}, \\
 \frac{\epsilon[1-P(x_i=m_i|m_1^{i-1})]}{P(x_i=m_i|m_1^{i-1})+\epsilon[1-2P(x_i=m_i|m_1^{i-1})]} & \quad \textrm{if $x_i=-m_i$}. 
\end{array}\right.\,
\label{choicehistory}
\end{equation}
Depending on the past measurements the probability $P(x_i=m_i|m_1^{i-1})$ on the right side of this equation
is either $p_t^{(i-1)}$ or $1-p_t^{(i-1)}$, where $p_t^{(i-1)}$ is obtained from $p_0^{(i-1)}$ and equation (\ref{pt}).
From equation (\ref{defp0i}), the state indicated by the measurement as occupied $m_i$ is lifted provided  $P(x_i=m_i|m_1^{i})<P(x_i=-m_i|m_1^{i})$,
which from (\ref{choicehistory}) is equivalent to $P(x_i=m_i|m_1^{i-1})<\epsilon$. Since $1-p_t^{(i-1)}\ge1/2$, a necessary condition for lifting the state $m_i$ at the beginning of period $i$ is that $p_t^{(i-1)}<\epsilon$. 

It is convenient to define the variables $z_i\equiv m_im_{i-1}$, for $i>1$, and $\sigma_i$, which takes the
value  $1$ ($-1$) if the level $-m_i$ ($m_i$) is lifted at the beginning of period $i$, i.e.,
\begin{equation}
\sigma_i \equiv \left\{
\begin{array}{ll} 
 1 & \quad \textrm{if $P(x_i=m_i|m_1^{i})>P(x_i=-m_i|m_1^{i})$}, \\
 -1 & \quad \textrm{if $P(x_i=-m_i|m_1^{i})>P(x_i=m_i|m_1^{i})$}.
\end{array}\right.\,
\end{equation}
Furthermore, we define $\tilde{z}_i\equiv z_i\sigma_{i-1}$. This last variable identifies whether for given $m_i$ the probability $P(x_i=m_i|m_1^{i-1})$ is
$p_t^{(i-1)}$ or $1-p_t^{(i-1)}$: 
\begin{equation}
P(x_i=m_i|m_1^{i-1})= \left\{
\begin{array}{ll} 
 1-p_t^{(i-1)} & \quad \textrm{if $\tilde{z}_i=1$}, \\
 p_t^{(i-1)} & \quad \textrm{if $\tilde{z}_i=-1$}.
\end{array}\right.\,
\label{pmx}
\end{equation}
In words, this equation means that if $\tilde{z}_i=1$ ($\tilde{z}_i=-1$) then $m_i$ corresponds to the lower (upper) level of period $i-1$.

For the initial condition before the first period we assume that the states are equally probable,
hence,  $p_0^{(1)}=\epsilon$ and $\sigma_1=1$. Numerical simulations of the measurement trajectory are then performed with the following algorithm:
\begin{enumerate}
\item[1)] For period $i$ randomly choose a measurement  according to the probability $P(m_i|m_1^{i-1})$ given by equations (\ref{errorh}) and (\ref{pmx});   
\item[2)] with $p_t^{(i-1)}$, the variable $\tilde{z}_i$, and equations (\ref{defp0i}), (\ref{choicehistory}) and (\ref{pmx}) calculate $\sigma_i$ and $p_0^{(i)}$;  
\item[3)] from relation (\ref{pt}) and  $p_0^{(i)}$ calculate $p_t^{(i)}$. Go back to the first step with the substitution $i\to i+1$. 
\end{enumerate}

This algorithm can be translated into a recursion relation for the initial probability. 
Using (\ref{choicehistory}) and (\ref{pmx}), relation (\ref{defp0i}) becomes 
\begin{equation}
p_0^{(i)}=  \left\{
\begin{array}{ll} 
 F_+(p_0^{(i-1)}) & \quad \textrm{if $\tilde{z}_{i}=1$}, \\
 \textrm{min}\{F_-(p_0^{(i-1)}),1-F_-(p_0^{(i-1)})\} & \quad \textrm{if $\tilde{z}_{i}=-1$}, 
\end{array}\right.\,
\label{iteapp}
\end{equation}
where 
\begin{equation}
 F_+(p_0^{(i-1)})\equiv \frac{\epsilon p_t^{(i-1)}}{1-q_t^{(i-1)}} 
\end{equation}
and
\begin{equation}
F_-(p_0^{(i-1)})\equiv \frac{\epsilon(1-p_t^{(i-1)})}{q_t^{(i-1)}}, 
\end{equation}
with $q_t^{(i-1)}\equiv p_t^{(i-1)}+\epsilon(1-2p_t^{(i-1)})$. Note that the function $1-F_-(p_0^{(i-1)})$ is minimal when $F_-(p_0^{(i-1)})>1/2$, which implies $\epsilon>p_t^{(i-1)}$. Only in this case, the state measured as occupied $m_i$
can be lifted at the beginning of period $i$.

Moreover, from (\ref{errorh}), (\ref{pmx}), and (\ref{iteapp}), the average initial occupation probability conditioned on $m_1^{i-1}$ is 
\begin{equation}
\langle p_0^{(i)}\rangle_i\equiv \sum_{m_i} p_0^{(i)} P(m_i|m_1^{i-1})= \textrm{min}\{p_t^{(i-1)},\epsilon\}.
\label{defavgi}
\end{equation}
If the machine never lifts the level $m_i$, i.e.,  $F_-(p_0^{(i-1)})>1/2$ for all periods, the above average becomes $\langle p_0^{(i)}\rangle_i=\epsilon$.

\section{Extracted work for the memory-less machine}
\label{appb}

For the memory-less machine we denote the initial occupation probability at period $i$ by
$\tilde{p}_0^{(i)}$. The final occupation probability at period $i$ is $\tilde{p}_t^{(i)}$:
as the memory-less machine does not use the optimal protocol, $\tilde{p}_t^{(i)}$ is not obtained from 
(\ref{pt}) but rather it is the solution of the master equation (\ref{eq:masterequation}) for a given protocol $\tilde{E}_\tau$
and initial condition $\tilde{p}_0^{(i)}$.   

Another difference in relation to the optimized model considered in \ref{appa} is that the variable $\sigma_i$ is not necessary for the memory-less machine, since here $\sigma_i=1$ for all $i$.
Hence, for the memory-less machine, equation (\ref{pmx}) becomes
\begin{equation}
P(x_i=m_i|m_1^{i-1})= \left\{
\begin{array}{ll} 
 1-\tilde{p}_t^{(i-1)} & \quad \textrm{if $z_i=1$}, \\
 \tilde{p}_t^{(i-1)} & \quad \textrm{if $z_i=-1$}.
\end{array}\right.\,
\label{pmx2}
\end{equation}
The iterative relation (\ref{iteapp}) then simplifies to
\begin{equation}
\tilde{p}_0^{(i)}=  \left\{
\begin{array}{ll} 
 \tilde{F}_+(\tilde{p}_0^{(i-1)}) & \quad \textrm{if $z_{i}=1$}, \\
 \tilde{F}_-(\tilde{p}_0^{(i-1)}) & \quad \textrm{if $z_{i}=-1$}, 
\end{array}\right.\,
\end{equation}
where 
\begin{equation}
 \tilde{F}_+(\tilde{p}_0^{(i-1)})\equiv \frac{\epsilon \tilde{p}_t^{(i-1)}}{1-\tilde{q}_t^{(i-1)}} 
\end{equation}
and
\begin{equation}
\tilde{F}_-(\tilde{p}_0^{(i-1)})\equiv \frac{\epsilon(1-\tilde{p}_t^{(i-1)})}{\tilde{q}_t^{(i-1)}}, 
\end{equation}
with 
\begin{equation}
\tilde{q}_t^{(i-1)}\equiv \tilde{p}_t^{(i-1)}+\epsilon(1-2\tilde{p}_t^{(i-1)}).
\label{qtilde} 
\end{equation}

Similar to (\ref{defavgi}), the average initial probability for fixed measurement history is
\begin{equation}
\langle \tilde{p}_0^{(i)}\rangle_i\equiv \sum_{m_i} \tilde{p}_0^{(i)} P(m_i|m_1^{i-1})= \epsilon.
\label{avgdum}
\end{equation}
We denote by $\tilde{W}_t(\tilde{p}_0^{(i)})$ the work that is obtained from (\ref{eq:work_in}) with 
protocol $\tilde{E}_\tau$ and initial probability $\tilde{p}_0^{(i)}$. The average work is then given by
\begin{equation}
\langle \tilde{W}_t(\tilde{p}_0^{(i)})\rangle_i= \tilde{W}_t(\langle\tilde{p}_0^{(i)}\rangle_i)= \tilde{W}_t(\epsilon),
\end{equation}
where in the first equality we used the fact that $\tilde{W}_t(\tilde{p}_0^{(i)})$ is linear in $\tilde{p}_0^{(i)}$. Since, $\tilde{W}_t(\epsilon)$ is independent of the history $m_1^{i-1}$, it follows that the average work is simply 
$\tilde{W}_t(\epsilon)$. For the memory-less machine we compare with the optimized one $\tilde{E}_\tau= E_\tau(\epsilon)$, which is given by (\ref{protopt}),  and the average work is $W_t(\epsilon)$, which is given by (\ref{maxwork}). Moreover, 
assuming that before the first measurement both states are equally probable, which leads to  $\tilde{p}_0^{(1)}=\epsilon$, the average work equals the work extracted in the first period.

\section{Critical line}
\label{appc}

\begin{figure}
\centering\includegraphics{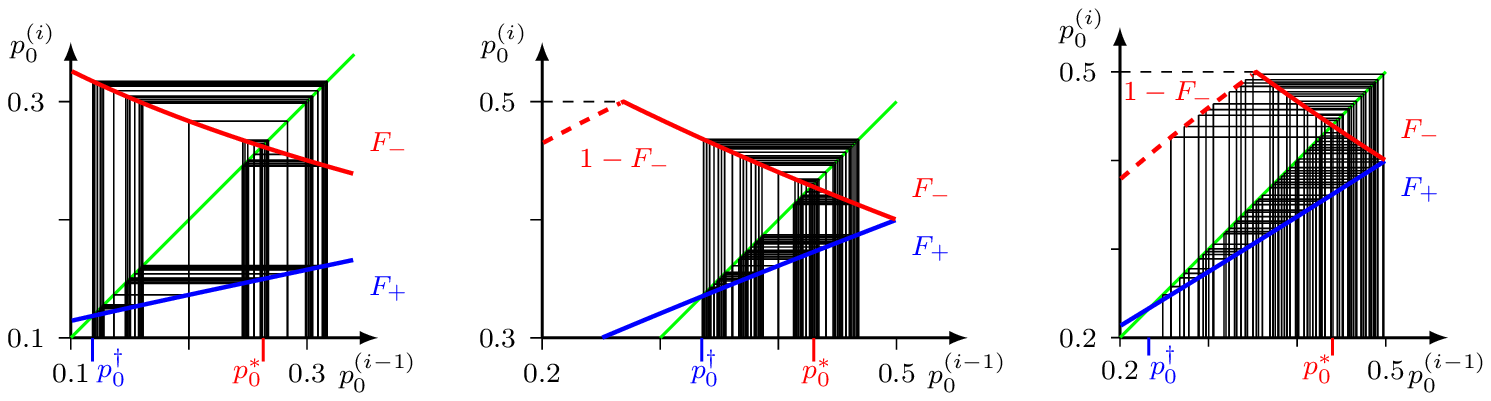}
\caption{Cobweb diagram of all possible first five iterations, starting with $p_0^{(1)}=\epsilon$, of the relation (\ref{recursion}).
In the left panel $t=2$ and $\epsilon=0.2$, in the central panel $t=1.5$ and $\epsilon=0.4$, and in the right panel $t=0.5$ and $\epsilon=0.4$.
For the first two cases $\phi=0$, whereas in the third case $\phi>0$.}
\label{fig4}
\end{figure}

We now obtain the critical line exactly by  analyzing the iterative relation for $p_0^{(i)}$ (\ref{recursion}). As discussed in \ref{appa}, the level $m_i$ will be lifted at the beginning of interval $i$
only if  $F_-(p_0^{(i-1)})>1/2$. Hence, in the phase $\phi=0$ the condition $F_-(p_0^{(i-1)})<1/2$ must be fulfilled.
The fixed points of these nonlinear maps are obtained from $F_+(p_0^\dagger)=p_0^\dagger$ and $F_-(p^*_0)= p^*_0$.
The possible trajectories in the cobweb diagram for the first five iterations of relation (\ref{recursion}) are shown in Fig. \ref{fig4}. It is clear that $p_0^{(i)}$ does not go below the fixed point $p_0^{\dagger}$.
Therefore, the critical line $\epsilon_c(t)$ can be obtained analytically  by setting
\begin{equation}
F_-(p^\dagger_0)= 1/2,
\label{eqcrit}
\end{equation}
which leads to a cumbersome transcendental equation.  Solving this equation we obtain the full black line in Fig \ref{fig3}, in perfect agreement with the numerical simulations.

Moreover, the phase $\epsilon<\epsilon_c$ can be further separated by two distinct regions, where the line (dotted line in Fig. \ref{fig3}) separating these two regions is obtained from 
\begin{equation}
F_-(0)=1/2.
\label{eqdot}
\end{equation}
In the region closer to the critical line with $F_-(0)>1/2$ the machine does not lift the state measured as occupied because $p_0^\dagger$ is not small enough, i.e., $F_-(p_0^\dagger)<1/2$ (see Fig. \ref{fig4}). However,
in this region depending on the initial condition the machine might lift the state measured as occupied during some initial transient.

\end{document}